\input{aipcheck}

\documentclass[
    ,final            
  ]
  {aipproc}

\layoutstyle{6x9}
\usepackage{graphics}
\usepackage{amssymb}
\usepackage{amsfonts}
\usepackage{amsmath}
\newcommand{\be}{\begin{eqnarray}}
 \newcommand{\ee}{\end{eqnarray}}
 \newcommand{\nee}{\nonumber\end{eqnarray}}
 \newcommand{\nn}{\nonumber\\ }

\begin{document}

\title{On Non-singlets in Kaon Production in  Semi-inclusive DIS reactions}


\author{Ekaterina Christova}{
  address={Institute for Nuclear Research and Nuclear Energy, Sofia, echristo@inrne.bas.bg}
}
\author{Elliot Leader}{
  address={Imperial College, London, UK, e.leader@imperial.ac.uk}
}

\begin{abstract}
 We consider semi-inclusive unpolarized DIS for the
production of charged kaons and the different possibilities, both
in LO and NLO, to test the conventionally used assumptions $s-\bar
s=0$ and $D_d^{K^+-K^-}=0$. The considered tests have the
advantage that they do not require
  any knowledge of the fragmentation functions.
\end{abstract}

\maketitle


\section{Introduction}

  It is well known that inclusive deep inelastic
scattering (DIS) yields information only about quark plus
antiquark parton densities. Semi-inclusive DIS (SIDIS), where some
final hadron $h$ is detected, plays an essential role to obtain
separate knowledge about quark and antiquark densities,  but it
requires a knowledge of the fragmentation function (FF) for a
given parton to fragment into the relevant hadron. As pointed out
in \cite{Kretzer_we} and more recently in  \cite{deFlorian} a
precise knowledge of the FFs is vital.

When the spin state of the detected hadron is not monitored, it is
possible to learn about the FFs from, both, $e^+e^-\rightarrow hX $
and \textit{unpolarized} SIDIS $ l+N \rightarrow l hX $. In the
case of pion production $SU(2)$ plays a very helpful role in
reducing the number of independent FFs needed. For charged kaon
production, which is important for studying the strange quark
densities, $SU(2)$ is less helpful, and even a combined analysis
of $e^+e^-$ and SIDIS data on both protons and neutrons does not
allow an unambiguous determination of the kaon FFs~\cite{Dubna05}.

It is thus conventional to make certain reasonable sounding
assumptions about the strange quark densities and the kaon FFs.
The usually made assumptions in analyzing the data are
$s(x) =\bar s(x) $ and $ D_d^{K^+}(z) =D_d^{K^-}(z)$. In
this paper we discuss to what extend these assumptions can be
justified and tested experimentally in, both, LO and NLO in QCD.  We
examine possible tests for the reliability of a leading order (LO)
treatment of the considered processes.

\section{Unpolarized  SIDIS $e+N\to e+h+X$ }

 In semi-inclusive deep inelastic scattering we consider the non-singlet
 difference of cross-sections
 $\tilde\sigma_N^{h-\bar h}$. The measurable quantity is the ratio
 of semi-inclusive $\tilde\sigma_N^{h-\bar h}$ and inclusive $\tilde\sigma_N^{DIS}$
 deep inelastic lepton-nucleon scattering:
 \be
R_N^{h-\bar h}=\frac{ \tilde\sigma_N^{h-\bar h}}{\tilde\sigma_N^{DIS}},\qquad
\tilde\sigma_N^{h-\bar h}= \tilde\sigma_N^{h}-\tilde\sigma_N^{\bar h}.
\ee
 For simplicity, we use  $\tilde\sigma_N^h$ and
$\tilde\sigma_N^{DIS}$ in which common kinematic factors have been
removed~\cite{strategy}. For $\tilde\sigma_N^{DIS}$ any of the
parametrizations for the structure functions $F_2$ and $R$ or,
equivalently, any set of the unpolarized parton densities can
be used.

As shown in~\cite{Dubna05}, the general expression for the  cross
section differences in NLO, is:
 \be
 \quad\tilde \sigma_p^{h-\bar h}(x,z)=\frac{1}{9}\left[4  u_V\otimes
D_u^{h-\bar h} +  d_V\otimes D_d^{h-\bar h} + (s -\bar s)\otimes
D_s^{h-\bar h}\right]\otimes  \hat\sigma_{qq} (\gamma q \to q
X)\nn
 \quad \tilde \sigma_n^{h-\bar h}(x,z) =\frac{1}{9}\left[4
d_V\otimes D_u^{h-\bar h} +  u_V\otimes D_d^{h-\bar h} + (s -\bar
s)\otimes D_s^{h-\bar h}\right]\otimes  \hat\sigma_{qq} (\gamma q
\to q X). \label{diff}
 \ee
Here $\hat\sigma_{qq}$ is  the perturbatively  calculable, hard
partonic cross section $q\gamma^*\to q+X$:
\be
 \hat\sigma_{qq} &=&  \hat\sigma_{qq}^{(0)} +
 \frac{\alpha_s}{2\pi} \hat\sigma_{qq}^{(1)}\,,
 \ee
normalized so that $ \hat\sigma_{qq}^{(0)} = 1 $, $D_q^{h-\bar h} \equiv D_q^h-D_q^{\bar h}$.

It is seen that  $ \tilde \sigma_N^{h-\bar h}$ involves only  NS
parton densities and fragmentation functions, implying that its
$Q^2$ evolution is relatively simple.
 Eq.(\ref{diff})  is sensitive to the valence quark
densities,  but also to the completely unknown combination
$(s-\bar s)$. The term $(s-\bar s)D_s^{h-\bar h}$ plays no role in
pion production, since, by SU(2) invariance,
$D_s^{\pi^+-\pi^-}=0$. However it is important for kaon
production, for which $D_s^{K^+-K^-}$ is a favoured transition,
and thus expected to be big.

Up to now all analyses of experimental data  assume $s=\bar
s$. In the next Sections we shall consider  the production of
charged kaons, $h=K^\pm$  and show how this
assumption,  and the assumption $D_d^{K^+-K^-}=0$, can be tested
without requiring knowledge of the FFs.

SU(2) symmetry is of little help if only charged kaons are measured.
However, it is well known that charged and neutral kaons are combined into SU(2) doublets.
This relates the FFs of $K_s^0$ to those of $K^\pm$, which implies that no new FFs appear in $K_s^0$-production.
In  ~\cite{kaons_we}  we examine to what extend detecting neutral  as well as charged
kaons can help to determine the kaon fragmentation functions. We carry out the analysis in LO and NLO.
 We show that  the non-singlet combination $(D_u-D_d)^{K^++K^-}$
can be measured directly both in $e^+e^-$ and in SIDIS without any influence of the strange and gluon parton densities
or any other FFs and this allows
 tests of the factorization of SIDIS
into parton densities and fragmentation functions in any order in QCD.

\section{Production of charged kaons }


As seen from (\ref{diff}), in $R_N^{K^+-K^-}$ both $s-\bar s$ and
$D_d^{K^+-K^-}$ appear. They are expected to be small, and the
usual assumption is that they are equal to zero. Here we examine
to what extent one can test these assumptions experimentally in
SIDIS.

 Up
to now, all analysis of experimental data have been performed
assuming both $s=\bar s$ and $D_d^{K^+-K^-}=0$.
 Note, that from the quark content of $K^\pm$, the assumption
$D_d^{K^+-K^-}=0$ seems very reasonable if the $K^\pm$ are
directly produced. However, if they are partly produced via
resonance decay this argument is less persuasive.

\subsection{ LO approximation}

 In LO we have:
\be
\tilde\sigma_p^{K^+-K^-} &=& \frac{1}{9} [4\,u_V\,D_u^{K^+-K^-} +
d_V\,D_d^{K^+-K^-} +(s-\bar s)\,D_s^{K^+-K^-}],\\
\tilde\sigma_n^{K^+-K^-} &=& \frac{1}{9} [4\,d_V\,D_u^{K^+-K^-} +
u_V\,D_d^{K^+-K^-} +(s-\bar s)\,D_s^{K^+-K^-}].
\ee

From a theoretical point of view it is more useful to consider the
following combinations of cross-sections, which, despite involving
 differences of cross-sections, are likely to be large:
 \be
 (\tilde\sigma_p - \tilde\sigma_n)^{K^+-K^-}=\frac{1}{9} [ (u_V-d_V)\,(4D_u-D_d)^{K^+-K^-}]\\
 (\tilde\sigma_p + \tilde\sigma_n)^{K^+-K^-}=\frac{1}{9} [
(u_V+d_V)\,(4D_u+D_d)^{K^+-K^-}+2(s-\bar s)D_s^{K^+-K^-} ]
\ee
We define:
\be
R_+(x,z)\equiv\frac{(\tilde\sigma_p +
\tilde\sigma_n)^{K^+-K^-}}{u_V+d_V},\qquad R_-(x,z)\equiv\frac{(\tilde\sigma_p -
\tilde\sigma_n)^{K^+-K^-}}{u_V-d_V}.
\ee

From a study of the $x$ and $z$ dependence of these we can deduce
the following:

1) if  $R_-(x,z)$ is a function of $z$ only, then  this suggests
that a LO approximation is reasonable.

2) if $R_+(x,z)$ is  \textit{also} a function of $z$ only, then,
since $D_s^{K^+ - K^-} $ is a favoured transition, we can conclude
that $(s-\bar s)=0$.

 3) if $R_+(x,z)$ and $R_-(x,z)$ are \textit{both} functions of $z$ only, and if in addition,
 $R_+(x,z) = R_-(x,z)$, then both $s-\bar s=0$
{\it and} $D_d^{K^+-K^-}= 0$.

4) if $R_+(x,z)$ and $R_-(x,z)$ are \textit{both} functions of $z$
only, but they are \textit{not} equal, $R_+(x,z) \neq R_-(x,z)$, we conclude that $s-\bar
s=0$, {\it but} $D_d^{K^+-K^-}\neq 0$.

5) if $R_-(x,z)$ is not a function of $z$ only, then NLO corrections
are needed, which we  consider below.

The above tests for $s-\bar s =0$ and $D_d^{K^+-K^-}= 0$ can be
spoilt either by $s-\bar s \neq 0$ and/or $D_d^{K^+-K^-}\neq 0$,
or by NLO corrections, which are both complementary in size.
That's why it is important to formulate tests sensitive to $s-\bar
s =0$ and/or $D_d^{K^+-K^-}= 0$ solely, i.e. to consider NLO.

\subsection{NLO approximation}

If an NLO treatment is necessary it is still possible to reach
some conclusions, though less detailed than in the LO case. We now
have:
 \be
(\tilde\sigma_p - \tilde\sigma_n)^{K^+-K^-}&=& \frac{1}{9} (u_V-d_V)\,
\otimes\,(1+\alpha_s\,C_{qq})\otimes\,(4D_u-D_d)^{K^+-K^-}\label{NLOp}\\
(\tilde\sigma_p + \tilde\sigma_n)^{K^+-K^-}&=& \frac{1}{9} \left[(u_V+d_V)\,\otimes\,(4D_u+D_d)^{K^+-K^-}\right.\nn
&&\quad\left.+2(s-\bar s)\,\otimes\,D_s^{K^+-K^-}\right]\,\otimes\,(1+\alpha_s\,C_{qq})\label{NLOn}
\ee

Suppose we try to fit  both (\ref{NLOp}) and (\ref{NLOn}) with one
and the same fragmentation function $D(z)$:
\be
&&
(\tilde\sigma_p - \tilde\sigma_n)^{K^+-K^-}\approx \frac{4}{9} (u_V-d_V)\,\otimes\,(1+\alpha_s\,{\cal C}_{qq})\otimes\,D(z),\\
&&(\tilde\sigma_p +
\tilde\sigma_n)^{K^+-K^-}\approx \frac{4}{9}
(u_V+d_V)\,\,\otimes\,(1+\alpha_s\,{\cal C}_{qq})\otimes\,D(z).
\ee

If this gives an acceptable fit for the $x$ and $z$-dependence of
both $p-n$ and $p+n$ data, we can conclude that both $s-\bar s
\approx 0$ {\it and} $D_d^{K^+-K^-}\approx 0$, and that $D(z) =
D_u^{K^+-K^-}$.

Note that for all above tests, both in LO and NLO approximation,
we don't require a knowledge of $D_{u,d}^{K^+-K^-}$. This is
especially important since the  $e^+e^-$ total cross section data
determine only the $D_{q}^{K^++K^-}$, and these are relatively
well known, while $D_{u,d}^{K^+-K^-}$ can be determined solely
from $A_{FB}$ in $e^+e^-$  or from SIDIS.

The results of the above tests would indicate what assumptions are
reliable in trying to extract the   fragmentation functions
$D_{u,d,s}^{K^\pm}$ from the same data.

\begin{theacknowledgments}
 This work was supported by a Royal Society International Joint Project Grant.
\end{theacknowledgments}

\end{document}